\documentclass[letterpaper,twocolumn,10pt]{article}
\usepackage{usenix2019_v3}

\usepackage{tikz}
\usepackage{array}
\usetikzlibrary{calc}
\newcommand{\tikzmark}[1]{\tikz[remember picture,overlay]\coordinate (#1);}
\newcolumntype{T}[1]{@{\hspace{\tabcolsep}}c@{\hspace{\tabcolsep}\tikzmark{#1}}}
\usepackage[T1]{fontenc}
\usepackage[utf8]{inputenc}
\usepackage{tcolorbox}
\usepackage{color}
\usepackage{graphicx}
\usepackage{framed}
\usepackage[ruled, vlined]{algorithm2e}
\usepackage{pifont}
\usepackage{standalone}
\usepackage{amsmath}
\usepackage{amsfonts}
\usepackage{amssymb}
\usepackage{paralist}
\usepackage{enumitem}
\usepackage{caption}
\usepackage{subcaption}
\usepackage{mdwlist}
\usepackage[mathcal]{euscript}
\usepackage{booktabs}
\usepackage{tabularx}
\usepackage[toc]{glossaries}
\usepackage{xspace}
\usepackage{balance,multirow}
\usepackage{microtype}
\microtypecontext{spacing=nonfrench}
\usepackage{pgf-umlsd}
\usepackage{hyperref}

\renewcommand{\paragraph}[1]{\medskip \noindent {\bf #1}.}
\long\def\symbolfootnote[#1]#2{\begingroup%
\def\thefootnote{\fnsymbol{footnote}}\footnote[#1]{#2}\endgroup}

\newcommand{\delphinium}{{Delphinium}}
\newcommand{\sat}{{SAT}}
\newcommand{\maxsharpsat}{{Max\#SAT}}
\newcommand{\sharpsat}{{\#SAT}}
\newcommand{\sharpp}{{\#P}}

\let\latexcite=\cite
\def\cite{\nolinebreak\latexcite}
\let\latexref=\ref
\def\ref{\nolinebreak\latexref}

\newcommand{\ignore}[1]{}

\newcounter{romanlistcounter}
  {\setcounter{romanlistcounter}{0}%
   \begin{list}{\textit{(\roman{romanlistcounter})}}{%
        \usecounter{romanlistcounter}%
      \setlength{\itemsep}{0pc}%
      \setlength{\itemindent}{1pc}%
      \setlength{\topsep}{0pc}%
      \setlength{\mylabelwidth}{3em}}}
  {\end{list}}

\newcounter{alphalistcounter}
  {\setcounter{alphalistcounter}{0}%
   \begin{list}{\textit{(\alph{alphalistcounter})}}{%
        \usecounter{alphalistcounter}%
      \setlength{\itemsep}{0pc}%
      \setlength{\itemindent}{1pc}%
      \setlength{\topsep}{0pc}%
      \setlength{\mylabelwidth}{3em}}}
  {\end{list}}

\def\compactify{\itemsep=0pt \topsep=0pt \partopsep=0pt \parsep=0pt}
\let\latexusecounter=\usecounter

\let\latexusecounter=\usecounter

\newcommand*\emptycirc[1][1ex]{%
    \begin{tikzpicture}[baseline=-\the\dimexpr\fontdimen22\textfont2\relax]
    \draw (0,0) circle (#1); 
    \end{tikzpicture}}
\newcommand*\dotcirc[1][1ex]{%
    \begin{tikzpicture}[baseline=-\the\dimexpr\fontdimen22\textfont2\relax]
    \draw[fill] (0,0) circle (#1/3) ;
    \draw (0,0) circle (#1);
    \end{tikzpicture}}
\newcommand*\halfcirc[1][1ex]{%
    \begin{tikzpicture}[baseline=-\the\dimexpr\fontdimen22\textfont2\relax]
    \draw[fill] (0,0)-- (90:#1) arc (90:270:#1) -- cycle ;
    \draw (0,0) circle (#1);
    \end{tikzpicture}}
\newcommand*\fullcirc[1][1ex]{%
    \begin{tikzpicture}[baseline=-\the\dimexpr\fontdimen22\textfont2\relax]
    \draw[fill] (0,0) circle (#1); 
    \end{tikzpicture}} 

\usepackage[
  separate-uncertainty = true,
  multi-part-units = repeat
]{siunitx}
\usepackage{listings}
\definecolor{dkgreen}{rgb}{0,0.6,0}
\definecolor{gray}{rgb}{0.5,0.5,0.5}
\definecolor{mauve}{rgb}{0.58,0,0.82}

\usepackage{float}
\usepackage{chngpage}
\usepackage{adjustbox}

\newcommand{\mcmpc}{{McFIL}}

\begin{document}

\date{}
\title{McFIL: Model Counting Functionality-Inherent Leakage\thanks{To appear at USENIX Security 2023 and accepted for publication in the Winter cycle of the proceedings thereof.}}

\author{
{\rm Maximilian Zinkus}\\
Johns Hopkins University \\
zinkus@cs.jhu.edu
\and
{\rm Yinzhi Cao}\\
Johns Hopkins University\\
yzcao@cs.jhu.edu
\and
{\rm Matthew D. Green}\\
Johns Hopkins University\\
mgreen@cs.jhu.edu
}

\maketitle

\begin{abstract}
Protecting the confidentiality of private data and using it for useful collaboration have long been at odds. Modern cryptography is bridging this gap through rapid growth in secure protocols such as multi-party computation, fully-homomorphic encryption, and zero-knowledge proofs. However, even with provable indistinguishability or zero-knowledgeness, confidentiality loss from leakage \textit{inherent to the functionality} may partially or even completely compromise secret values without ever falsifying proofs of security.

In this work, we describe \mcmpc, an algorithmic approach and accompanying software implementation which automatically quantifies intrinsic leakage for a given functionality. Extending and generalizing the Chosen-Ciphertext attack framework of Beck \textit{et al.} with a practical heuristic, our approach not only quantifies but \textit{maximizes} functionality-inherent leakage using Maximum Model Counting within a \sat~solver. As a result, \mcmpc~automatically derives approximately-optimal adversary inputs that, when used in secure protocols, maximize information leakage of private values.
\end{abstract}

\section{Introduction}

Functionality-inherent leakage (FIL) is a universal characteristic of systems which compute over private data. Modern cryptography has enabled many such systems, including secure multiparty computation (MPC), fully-homomorphic encryption (FHE)~\cite{gentry2009fully}, and zero-knowledge (ZK) proofs~\cite{ishai2007zero}. The use of these cryptographic tools is growing steadily across the public and private domains, increasing the stakes of these systems' security and increasingly bifurcating the set of people who develop these systems from the set who use and rely on them. 

FIL occurs when an adversary is able to observe or participate in computation over private data, and observe the outputs of this computation (or factors correlated with them). Naturally, given a computable function and even partial knowledge of some inputs and outputs, an adversary can infer \textit{some}thing about the unknown inputs which induced observed outputs.

In this work, we provide~\mcmpc, an algorithmic approach for evaluating FIL within arbitrary functionalities, and an accompanying software implementation which can be used to determine the extent of leakage a functionality admits. Our approach does not exploit any cryptographic insecurity; rather, we leverage the unavoidable leakage inherent to underlying functions. The fact that cryptographic protocols can reveal information via correctly-evaluated outputs should be no surprise to cryptographers.
It is our aim to provide a systematic way for prospective, non-expert users of cryptographic systems to evaluate functionalities they wish to compute securely, so they can make informed decisions on the risks of doing so even within cryptographically-secure schemes.

\begin{figure}[t]
    \centering
    \includegraphics[width=\linewidth]{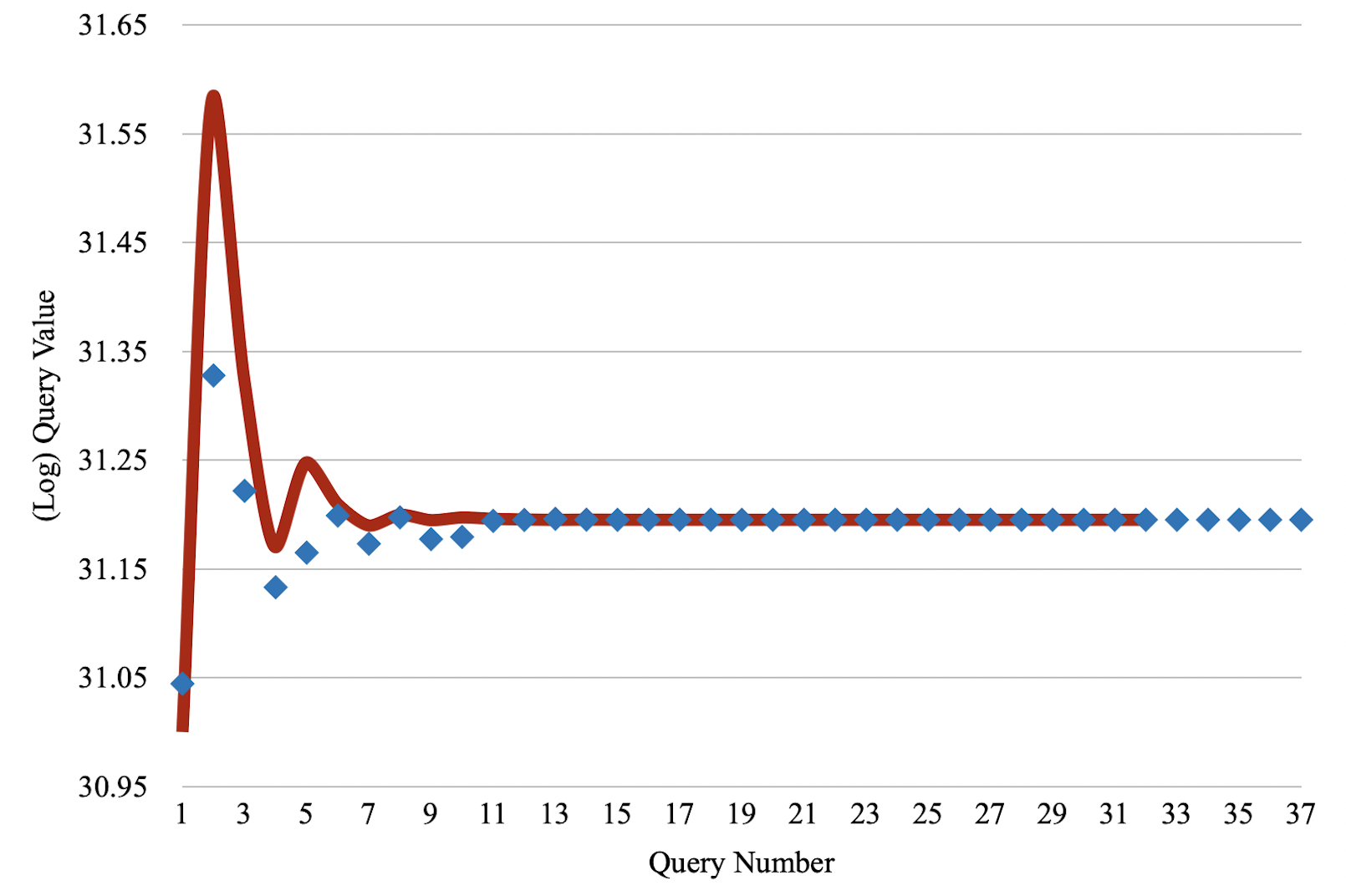}
    \caption{In a fully-automated analysis of Yao's Millionaires, \mcmpc~``discovers'' binary search. Blue diamonds are adversary inputs, true binary search shown in solid red.
    }
    \label{fig:millionaires_appmc}
\end{figure}

\paragraph{Our approach} We rely on a family of techniques called Model Counting, and we refer to our tool as \mcmpc~for ``Model Counting Functionality-Inherent Leakage.'' Most critically, \mcmpc~is designed to quantify and optimize leakage {\em given only a description of the circuit to be implemented}, and does not require the implementer to assist the tool in understanding the functionality. This allows the tool to automatically derive a number of ``attacks,'' including those not easily predicted by practitioners. For example:

\begin{itemize}
    \item The classic Yao's Millionaires problem~\cite{yao1982protocols} admits one bit of leakage per execution due to the functionality (greater-than comparison). Given only a description of the functionality, \mcmpc~automatically derives inputs to uncover the other player's salary in approximately $log(n)$ sequential executions (Figure~\ref{fig:millionaires_appmc}). 
    \item Dual Execution MPC~\cite{mohassel2006efficiency,huang2012quid} admits adversary-chosen one-bit leakage in an equality check protocol. \mcmpc~derives a sequence of predicates of configurable complexity to uncover the honest party's input.
    \item We evaluate \mcmpc~against an array of other functionalities as proofs of concept in \S\ref{sec:evaluation}, either completely recovering the honest parties input(s) or providing an equivalence class of candidate solutions many orders of magnitude smaller than the initial search space.
\end{itemize}

\paragraph{Leakage Explained} When considering novel functionalities for use in secure protocols such as MPC, FHE, or ZK, one must consider how the privacy of secret inputs will be maintained.
Therefore, practitioners must consider the security of their schema but also what can be inferred from the outputs of the functionalities. Even with provably secure protocols, the confidentiality loss inherent to a given functionality may partially or even completely leak secret values without ever violating the security guarantees of a protocol.

While functionality-based leakage may be unavoidable, it is quantifiable. Prior works~\cite{clarkson2006quantifying,mardziel2011dynamic,mardziel2013dynamic,mardziel2013knowledge} have applied various information flow and optimization techniques to this problem. However, quantifying information flow~\cite{clark2004quantified} over programs is \texttt{NP}-hard and grows in the complexity of the program state space~\cite{klebanov2013sat}. As a result, these approaches have remained computationally infeasible in practice. In addition to these general results, bespoke analyses for individual protocols or leakage paradigms exist in the literature~\cite{boyle2012multiparty,hazay2020going,almashaqbeh2021gage}, but these require manual analysis effort by experts which does not scale.

\paragraph{Contributions} We provide a practical methodology for \textit{automatically quantifying} and even \textit{optimizing over} inherent leakage of a circuit-based functionality. Our tool {\mcmpc} can be used to analyze privacy in MPC, FHE, or ZK. We generalize and extend a series of techniques developed by Beck \textit{et al.}~\cite{delphinium} in automating chosen-ciphertext attacks (CCA) in order to bring their work to a far broader class of functionalities.

\medskip \noindent
In summary, our contributions are as follows:
\begin{itemize}
  \setlength{\itemsep}{0em}
    \item We describe {\mcmpc}, an algorithmic approach to quantifying and exploiting information leakage for a given functionality. (\S\ref{sec:mcmpc})
    \item We generalize and extend the \delphinium~cryptanalysis framework of Beck {\em et a.}~\cite{delphinium} beyond its original domain of strictly-defined predicate functions, addressing key limitations of that work and resolving an open question from it.
    (\S\ref{sec:beyondpredicates})
    \item We provide a software implementation~\cite{mcmpc_code}, as an artifact accompanying this submission and open-source tool. Our implementation encodes nontrivial domain knowledge to directly manipulate \sat~instances and enable extensive parallelism. (\S\ref{sec:impl})
    \item Finally, we provide \sat~instances generated from \mcmpc~and our sample programs.
    The \sat~community gathers such instances as benchmarks~\cite{hoos2000satlib} to guide future \sat~solver development, which will in turn expand the range and complexity of functions our tool can analyze within a given amount of computation time. (\S\ref{sec:benchmarks})
\end{itemize}


\section{Intuition: Maximizing Leakage}\label{sec:intuition}

In the following two illustrative examples, we analyze simple functionalities to manually derive the results which \mcmpc~automates.

\subsection{A Minimal Example: Binary \textit{AND}}

Consider a simplistic two-party MPC (2PC) protocol which securely computes the functionality $\mathcal{F}$ consisting only of a Boolean $AND$ gate ($\land$).
\newcommand{\ztwo}{\mathbb{Z}_2}
\begin{quote}
$\ztwo = \{0, 1\}$\\
$\mathcal{F}: \ztwo \times \ztwo \to \ztwo = \land$\\
$a \in \ztwo$: honest party's input\\
$b \in \ztwo$: adversary's input\\
$x \in \ztwo$: output ($x = a \land b$)
\end{quote}

\paragraph{Maximizing Leakage} After normal execution, the adversary should have no knowledge of $a$. However, knowing $\mathcal{F} = \land$, the adversary may \textit{a priori} choose their input $b = 1$ to gain complete knowledge of $a$ upon learning $x$:
\begin{quote}
    $x = 0 \implies a \land 1 = 0 \implies a = 0$\\
    $x = 1 \implies a \land 1 = 1 \implies a = 1$
\end{quote}

The choice of $b = 1$ \textit{maximizes the information} gained from $x$. Had they chosen $b = 0$, $x = 1$ would have become impossible for the $AND$ functionality, and thus the output would provide no information about $a$.
By choosing $b$ optimally, the adversary learns $a$ despite the security of the MPC.

\subsection{A Further Example: Yao's Millionaires}

Now, consider the classic 2PC example of the Millionaires problem~\cite{yao1982protocols,yao86}. Two parties ($A$, $B$) wish to compare their wealth (say, as 32-bit integers $a$ and $b$) without disclosing anything aside from the predicate result of $a < b$.
\newcommand{\ztt}{\mathbb{Z}_{32}}
\begin{quote}
$\ztt = \{0, \ldots, 2^{32}-1\}$\\
$\mathcal{F}: \ztt \times \ztt \to \ztwo~=~<$\\
$a \in \ztt$: honest party's input\\
$b \in \ztt$: adversary's input\\
$x \in \ztwo$: output ($x = a < b$)
\end{quote}

\paragraph{Maximizing Leakage} In this example, no single adversary input will uniquely constrain the honest party's input (at least, not in expectation over a uniform distribution of possible $a$).

Now, consider $b = 2^{31} - 1$. If $x = 0$ (\texttt{false}), the most significant bit of the honest input must be $1$ as $a \geq 2^{31}$. Equivalently, if $x = 1$, then $MSB(a) = 0$. Assuming uniform $a$, in expectation the adversarial input $b = 2^{31} - 1$ eliminates half of the candidate solutions for $a$, providing the adversary a single bit of information.

\paragraph{Extending Leakage} On its own, this single bit of information out of 32 bits is potentially unimportant. The number of possible solutions for $a$ remains large, and the adversary has no way to distinguish between them.

Intuitively, we next consider: \textit{what if the adversary can try again?} Although this extends beyond the original threat model of many secure protocols, it does so with pragmatism: one can imagine settings where a protocol may be executed multiple times e.g. in a client-server model, when attempt-limiting protections are missing or evaded, or when an honest party may unwittingly interact with multiple colluding adversaries.

\paragraph{Maximizing \textit{Multi-Run Adaptive} Leakage} Given additional attempts (or ``queries,'' following~\cite{delphinium}), a binary search emerges from the Millionaires comparison functionality. When the protocol may be repeated, $a$ can be completely uncovered in $log(|\ztt|) = 32$ queries, violating confidentiality despite a secure protocol, and making \textit{optimally-efficient use} of the newly-assumed threat model of multiple queries.

This should be unsurprising to cryptographers: if security is proven modulo leakage, and leakage is allowed to grow, naturally security may be compromised. However, automatically analyzing and quantifying leakage over many queries is useful in two ways. First, the more intuitive: if a protocol does admit some method to retry, for example a smart contract~\cite{ethereum} which may be repeatedly executed or an online oracle which does not sufficiently authenticate users, a multi-run leakage amplification attack may be possible. In this case, \mcmpc~can be used to derive a set of queries to efficiently exploit leakage. Second, the multi-run setting can be useful in analyzing protocols which do not admit multiple attempts in that \mcmpc~can compute an average leakage statistic per query. For arbitrary protocols, the true extent of leakage may be unknown; \mcmpc~provides a way for practitioners to evaluate functionalities before choosing to encode them into secure protocols. As demonstrated in Figure~\ref{fig:millionaires_appmc}, \mcmpc~approximately rediscovers this binary search attack to uniquely identify the \texttt{target} secret input.

\begin{figure}[!t]
    \centering
    \includegraphics[width=0.6\linewidth]{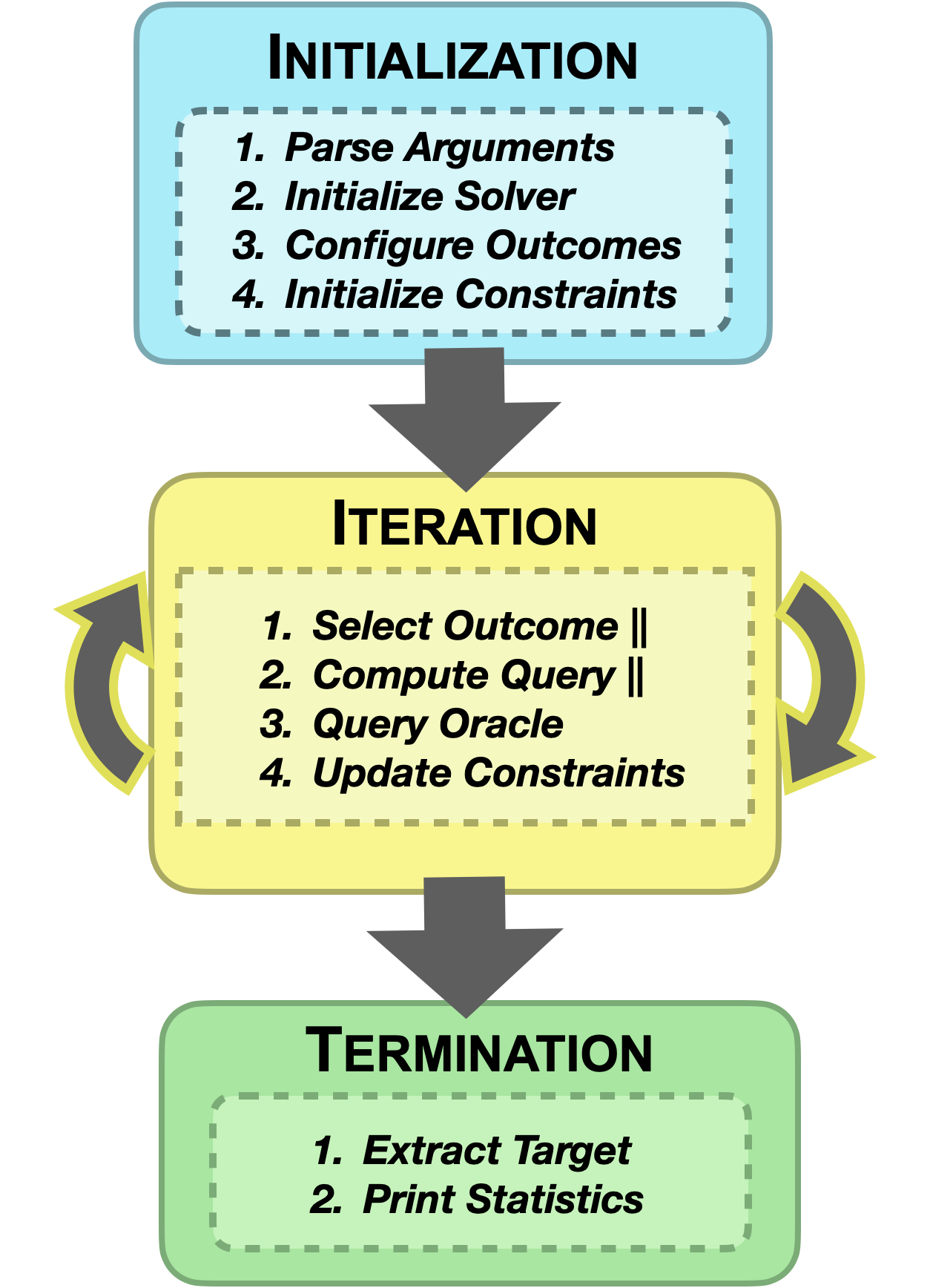}
    \caption{\mcmpc~workflow; $||$ denotes parallelization}\label{fig:mcmpc_workflow}
\end{figure}

\subsection{Overview of McFIL}
Here we provide intuition as to how \mcmpc~uses \sat~solving to achieve automated leakage maximization. Refer to Figure~\ref{fig:mcmpc_workflow} for a visualization of our automated workflow.

\paragraph{Prior Work: \delphinium} Beck \textit{et al.} applied \sat~solving to the problem of Chosen Ciphertext attacks (CCAs) in their tool, \delphinium~\cite{delphinium}. Their methodology formulates CCAs as a sequence of optimization problems over the outputs of a format oracle, automating attacks against classic and novel oracles to exploit CCA vulnerabilities.

Format oracles are predicate functions which determine if an input is well-formed or not, and \delphinium~exploits this leakage to decrypt. In this work, we follow the methodology of Beck \textit{et al.} and formulate generalized functional leakage as a similar sequence of optimization problems. The main contributions of this work, then, are the necessary adaptations to generalize their approach from predicates (1-bit outputs) to arbitrary $n$-bit output functions while maintaining concrete efficiency in real-world target functionalities. Further, \delphinium~requires its format functions to be well-defined (e.g. for all possible inputs, either $true$ or $false$ is returned). To accurately capture the breadth of arbitrary functions, we release this requirement using a novel, concretely-efficient heuristic described in \S\ref{sec:beyondpredicates}.

\paragraph{Initialization} To initialize \mcmpc, a Boolean circuit representation of a function functionality must be provided.
Compilers exist to facilitate this process~\cite{z3}, and \mcmpc~offers a Python3 DSL to facilitate encoding (refer to Appendix~\ref{app:dsl}). These instances contain symbolic bits representing the ``target'' secret input(s) to be uncovered (\texttt{target}), and symbolic bits representing the adversary's ``chosen'' input(s) (generated by \mcmpc) or ``queries'' (\texttt{chosen}). 

\paragraph{Iterative Solution Elimination} In order to uncover secret input(s) in an automated, multi-round analysis, \mcmpc~requires access to a concrete instantiation of the functionality. Whether implemented as a test shim for local execution (as we provide in our implementation) or e.g. a live instance of a secure protocol, this instantiation must compute the functionality given the adversary's query and the true secret input(s) of the honest parties, and provide the result back to \mcmpc. We refer to this as the \textit{function oracle}, and each iteration corresponds to e.g. a single MPC evaluation in the \textit{multi-run adaptive} threat model.

Using a \sat~solver, these two components alone are sufficient to iteratively constrain the search space of the secret input(s). However, the queries generated at each iteration would be arbitrarily drawn from the set of satisfying models within the solver -- degenerating to a brute-force guessing attack within an exponential search space. What remains is to optimize the \textit{profitability} (expected number of eliminated models~\cite{delphinium}) of each query.

\paragraph{Optimizing Queries} A truly optimal leakage maximization would consider all possible numbers and contents of queries. However, due to the underlying complexity of the problem, this approach fails in practice. Existing leakage analysis work in side channels~\cite{phan2017synth} attempts to analyze a sliding window of multiple query-like values, and reaches computational limits in ~8-bit secret domains. Therefore, we consider only maximizing at each iteration: a greedy-optimal approach.

In order to maximize profitability at each iteration, we employ \maxsharpsat~to \textit{simultaneously maximize} the number of satisfying solutions for \texttt{target} corresponding to all possible outputs of the function under test. \mcmpc~then iteratively derives function inputs which imply partitions of the solution space, and eliminates one of the subsets of each partition based on the result from the function oracle.

\maxsharpsat, described in detail in \S\ref{sec:background}, is a counting-maximization analogue to Boolean Satisfiability (\sat) which can be efficiently, probabilistically approximated~\cite{fremont-aaai17}. Crucially, these approximation algorithms can themselves be efficiently encoded into \sat~constraints~\cite{delphinium}. By using these approximations as constraints in \sat~instances, \mcmpc~probabilistically and approximately ensures that a large number of satisfying solutions exist for given symbolic variables.

\paragraph{Extracting a Model} With all constraints in place and an unknown \texttt{target}, a satisfying solution (or ``model'') for \texttt{chosen} is extracted from the solver. The maximization constraints ensure that \texttt{chosen} is selected approximately optimally: for each possible output of $\mathcal{F}(\texttt{chosen}, \texttt{target})$, many candidate models for \texttt{target} exist.

Given only a symbolic representation of \texttt{target}, $\mathcal{F}$ cannot be used to compute a concrete result. However, when the concrete protocol instantiation -- containing knowledge of the true value of \texttt{target} -- is executed with the adversary's \texttt{chosen} input, a single outcome \texttt{result} is returned. Then, all candidate solutions for which $\mathcal{F}(\texttt{chosen}, \texttt{target}) \neq \texttt{result}$ may be eliminated. Critically, due to the maximization constraints, this set of eliminated candidates will be large.

\paragraph{Eliminating Candidates} After each query, all candidate models for \texttt{target} \textit{inconsistent} with each $(\texttt{chosen}, \texttt{result})$ pair are eliminated. That is, they are no longer satisfying solutions for \texttt{target} in the evolved constraint system. This leads to a greedy algorithm in expectation, iteratively reducing the \texttt{target} search space in maximized increments.

\paragraph{Beyond Predicates} Problematically, the \delphinium~algorithm is restricted to predicate functions, and intrinsically requires that for all \texttt{chosen}, every possible output for $\mathcal{F}$ is reachable. We refer to this informally as \textit{completeness} in \S\ref{sec:mcmpc}:
\begin{align*}
    \forall \texttt{chosen}, \texttt{result}~\exists \texttt{target} \\
    s.t.~\mathcal{F}(\texttt{chosen}, \texttt{target}) = \texttt{result}
\end{align*}

This completeness restriction is often trivial for predicates. While this is sufficient for chosen-ciphertext attack discovery~\cite{delphinium}, as the relevant padding/format functions are generally predicates which determine message validity, arbitrary functionalities are not necessarily so amenable. Further, \delphinium~requires the \textit{operator} to carefully define Boolean formulae to adhere to this notion of completeness, requiring additional effort and expert insight. We describe this challenge and our approach to generalize and extend \delphinium~in~\S\ref{sec:beyondpredicates}.

\paragraph{Negative Results: Demonstrating Security}
Depending on the functionality, it may be impossible to differentiate any classes of candidates; \mcmpc~detects this and provides a message that the attack may proceed as ``brute-force.'' This negative result can also be taken as an indication that leakage is bounded for the given functionality. If little enough of the private input is derived before brute-force is required, this may be taken as a probabilistic argument for the security of the functionality against leakage-based attacks.

\section{Technical Background}\label{sec:background}

In this section, we discuss \sat, its extensions, and their relative complexity. We also highlight software tools solving or approximating these problems which have emerged from the \sat~research community, and describe their relevance to \mcmpc. Finally, we review the limited past works which have broached function-inherent leakage and faced computational-feasibility limitations.

\subsection{SAT and SMT}

\paragraph{Boolean Satisfiability} \sat~is a widely-known NP-complete problem. \sat~takes as input a Boolean formula which relates a set of binary input values through Boolean operations. A solution to \sat~is an assignment (also called a model, solution, witness, or mapping) of \texttt{true} and \texttt{false} (eq. 1 and 0) values to the Boolean inputs which causes the formula to evaluate to \texttt{true}. \sat~formulae are commonly organized to include only $And$, $Or$, and $Not$ operations in Conjunctive Normal Form (CNF)~\cite{tseitin1968complexity}. Any propositional formula can be efficiently converted into CNF preserving satisfiability, although conversion may introduce a linear increase in formula size~\cite{tseitin1968complexity}. In CNF, \textit{literals} represent binary input values, and may be negated (denoted with $-$). These literals (negated or otherwise) are grouped into disjunctions ($Or$), which are themselves grouped into a single conjunction ($And$). In short, CNF is an ``and of ors.''


\paragraph{SAT Solvers} \sat~solvers are tools designed to determine the (un)satisfiability of Boolean formulae. \sat~solvers often require CNF, and provide a \textit{model} in the event the formula is satisfiable. The model is not guaranteed to be unique (and often is not).

Due to the generality of NP-complete problems~\cite{cook1971complexity}, \sat~solvers are powerful tools. Since 1962, the DPLL~\cite{davis1962machine} method of backtracking search has served as the core tool in \sat~solving, with the more recent development of Conflict-Driven Clause Learning (CDCL)~\cite{silva1996grasp} in 1996 aiding in optimizing the search for a satisfying model or contradiction indicating UNSAT.

\paragraph{Satisfiability Modulo Theories} Given a \sat~solver, the task remains to translate problems of interest into Boolean formulae. In order to aid translation, SMT solvers were developed. SMT solvers add expressive domains of constraint programming to \sat~such as arithmetic and bitvector (bitwise operations beyond single Boolean values) logic. SMT solvers such as Z3~\cite{z3} have enabled solving problems ranging from program verification~\cite{klee,angr,sage} to type inference and modeling, and more.

\subsection{Extensions to SAT}

Despite the expressive power of SMT solvers, program analysis alone is insufficient for \mcmpc~as we derive optimizations over the target circuit. As we \textit{maximize} over a solution space of program results, \mcmpc~generates instances of \maxsharpsat~(described in this section) from the Boolean formula describing the functionality under test. Intuitively, \maxsharpsat~is a strictly more complex problem than \sat, but can be approximated using the recent technique of Fremont \textit{et al.}~\cite{fremont-aaai17} and a \sat~solver. In this section we provide technical background describing \sat~through \maxsharpsat.

\begin{algorithm}[ht]
    \SetAlgorithmName{Definition}{}\\
    \KwIn{$\phi$: Boolean formula; \\\hspace{2.5em} $\overline{X}, \overline{Y}, \overline{Z}$: Vectors of Boolean inputs to $\phi$}
    \KwOut{$\overline{X}_{max}$: Assignment (concrete Boolean values) of variables in $\overline{X}$ for which the number of satisfying solutions to variables in $\overline{Y}$ is maximized and at least one satisfying solution exists in $\overline{Z}$; \\ \hspace{3.5em} $max$: Number of satisfying solutions for $\overline{Y}$ \\\hspace{3.5em} ($0 \leq max \leq 2^{|\overline{Y}|})$}
\caption{$\maxsharpsat(\phi, \protect\overline{X}, \protect\overline{Y}, \protect\overline{Z}) = \protect\overline{X}_{max}, max$}
\label{alg:maxsharpsat}
\end{algorithm}

\paragraph{SAT to \#SAT} \sharpsat, pronounced ``sharp~\sat,'' is the counting analogue to \sat. Also referred to as \textit{model counting}, \sharpsat~asks not only \textit{if} a satisfying model exists, but \textit{how many} such models exist. The result lies in a range from $0$ for an unsatisfiable formula, to $2^n$ for a (completely unconstrained) formula over $n$ bits. \sharpsat~is \sharpp-complete, at least as difficult as the corresponding NP problem, but this phrasing belies its complexity: Toda demonstrated PH $\subseteq$ P$^{\#P}$~\cite{Toda91}, that a polynomial-time algorithm able to make a single query to a \sharpp~oracle can solve any problem in PH, the entire polynomial hierarchy (which contains both NP and co-NP)~\cite{STOCKMEYER19761}.

\paragraph{\#SAT to Max\#SAT} \maxsharpsat~(denoted in Definition~\ref{alg:maxsharpsat}) is the optimization analogue to the \sharpsat~counting problem. As defined by Fremont \textit{et al.}, \maxsharpsat~takes a Boolean formula, denoted $\phi$, over Boolean variables divided into three notional subsets $\overline{X},\overline{Y}, \overline{Z}$. A solution to \maxsharpsat~provides a model for the variables in $\overline{X}$ such that the count (number of models) of the variables in $\overline{Y}$ is maximized and at least one model exists for the variables in $\overline{Z}$. ($\overline{Z}$ exists to allow variables to remain in the \sat~instance, yet outside the optimization constraints). A \maxsharpsat~solution is particularly useful when $\overline{X}$ refers to input variables to the formula, meaning that a model for $\overline{X}$ could be provided as an input to a program represented by the circuit $\phi$. Correspondingly, $\overline{Y}$ should be configured to contain variables of interest for maximization such as those representing quantitative information leakage, probabilistic inferences, or program-synthesis values~\cite{fremont-aaai17}.

\paragraph{Approximating Max\#SAT} \maxsharpsat~is at least as complex as \sharpsat~\cite{fremont-aaai17,Toda91}, but as Fremont \textit{et al.} demonstrates, this does not prevent approximating \maxsharpsat~using a number of calls to an NP oracle -- or in concrete terms, a \sat~solver. To approximate \maxsharpsat, Fremont \textit{et al.} rely on a sampling technique first described by Valiant~\cite{valiant79}, and expanded upon (and implemented in software) by Chakraborty and Meel \textit{et al.}~\cite{chakraborty2016algorithmic} and Soos~\cite{approxmc}. By applying \textit{almost-uniform} hash functions~\cite{valiant1979complexity,approxmc}, which representatively sample a domain with error, a non-uniform search space can be proportionately sampled with bounded error to enable a more feasible count. Crucially, these hashes can be efficiently sampled and applied within a \sat~solver~\cite{chakraborty2014distributionaware}. This approach is central to \mcmpc: using \textit{almost-uniform} hash functions as constraints within the formula, large classes of models for \texttt{target} can be identified and iteratively eliminated.

\subsection{Prior Work}

Functionality-based leakage is a known problem, although practical quantification methods are lacking in the literature. Clarkson \textit{et al.}~\cite{clarkson2006quantifying} analyze adversary knowledge of private MPC inputs using quantified information flow, but provide only a theoretical treatment of the problem. This idea was later explored by Mardziel \textit{et al.}~\cite{mardziel2011dynamic,mardziel2013knowledge,mardziel2013dynamic} in their attempt to maximize adversary knowledge through probabilistic polyhedral optimization. Although their solution is theoretically robust, Mardziel \textit{et al.} note the ``prohibitive'' computational cost of their approach in practice~\cite{mardziel2013knowledge}.

\sat~solving has also been applied to MPC in the recent literature, however, with particular focus on the \textit{intermediate values} generated during protocol execution. These intermediate values may leak some degree of information, which has been quantified using information flow~\cite{rastogi2013knowledge} and language-based formalization methods~\cite{almeida2018enforcing}. Further, when intermediate values can be determined to leak only \textit{negligible} information, prior work has shown they may be computed ``in-the-clear'' as an optimization~\cite{rastogi2013knowledge}.

Quantitative information flow, program analysis, and other automated approaches have also been applied in the detection and mitigation of side-channel leakage vulnerabilities. These vulnerabilities are similar to function-inherent leakage in that they provide additional signal which may compromise private inputs based on the (otherwise secure) execution of a function. A recent systematization~\cite{buhan2022sok} of side-channel detection literature enumerates various techniques ranging from micro-architectural modeling or even reverse engineering to program-analysis-like tools to analyze hardware descriptor languages for potential side channels. One such work, SLEAK~\cite{walters2014sleak}, computes a statistical distance between secret values and active intermediate values in the processes using a full-system simulator. The information-theoretic metric of leakage they compute is likely correlated with the leakage which \mcmpc~is able to uncover and maximize, however, their approach requires hardware simulation and a compiled binary, neither of which may be available or even relevant to target functionalities intended to be executed e.g. within an MPC.

\section{McFIL}\label{sec:mcmpc}

In this section, we describe our primary contribution, \mcmpc. We introduce the novel \texttt{SelectOutcomes} prioritization subroutine which extends the \maxsharpsat-based approach beyond \textit{complete} predicate functions and enables its use in the domain secure protocols such as MPC, FHE, and ZK.

\paragraph{Attack Model}
An ideal approach to identifying leakage would allow for the quantification of useful leakage after one round of execution. For novel function circuits, this analysis can inform operational security requirements and privacy considerations. However, in many cases a functionality will be executed multiple times on identical (or related) inputs.\footnote{This can occur in some protocols by design. Alternatively it may occur through deception, corruption of honest parties, or a failure of access control.} Ensuring the safety of all private inputs in these cases can also be seen as a form of defense-in-depth. For this setting, a \textit{multi-run adaptive} model allows {\mcmpc}~to extract more information about the privacy implications of a given Boolean circuit, whether the goal is to evaluate the safety of allowing untrusted parties to execute the protocol repeatedly (to automate an attack), or simply to to quantify and bound overall leakage.

\begin{figure}[!t]
    \centering
    \includegraphics[width=0.85\linewidth]{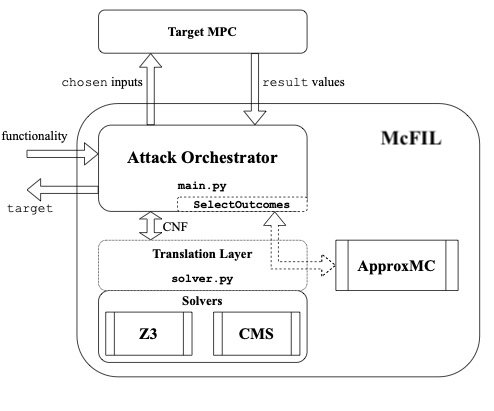}
    \caption{Architecture of \mcmpc}
    \label{fig:arch_diagram}
\end{figure}
\paragraph{System Architecture}
Figure~\ref{fig:arch_diagram} depicts the architecture of \mcmpc. Our tool integrates a custom CNF manipulation toolkit (\texttt{solver.py}) which efficiently represents constraints and orchestrates parallel solving instances.

\subsection{Beyond Predicate Functions}\label{sec:beyondpredicates}

In \delphinium~\cite{delphinium}, a format oracle classifies inputs as either valid or invalid according to a format specification and a predetermined fixed input length $l$. This requires format specifications to be \textit{complete} and \textit{deterministic}. Completeness requires that for every possible bit string of length $l$, the format function returns exactly one of \texttt{true} or \texttt{false}. Determinism requires that any input to the format function will always be classified the same way. As a result, the format function defines a two-set partition of the space of bit strings of length $l$, with the two sets corresponding to \texttt{true} and \texttt{false} under the format, respectively. Taken together, the requirements of \delphinium~ensure that at each iteration, the \sat~formula will remain satisfiable as long as a query exists which can differentiate at least some \texttt{true} and \texttt{false} partition elements.

\paragraph{Challenge Intuition} Clearly, arbitrary functions may be more complex than predicates. To address this, a natural first idea is to implement a third possible response (in addition to \texttt{true} and \texttt{false}), e.g. \texttt{error}. By introducing an ``error'' class, outputs could be simplified and the completeness requirement removed. However, adding a third output class raises a unique challenge which the \delphinium~authors leave to future work.

\mcmpc~seeks to \textit{simultaneously maximize} the number of secret inputs which correspond with \textit{all} $n$ output classes, such that each round, up to $\approx \frac{1}{n}$ of candidates may be eliminated in expectation. However, with the introduction of an error class and relaxed function definitions, the guarantee of simultaneous satisfiability is lost. For example, it may be that an input exists where some candidates would evaluate to a given output, and the rest to \texttt{error}, but none to another output. \delphinium~requires simultaneous satisfiability of all classes, and so the resulting \texttt{UNSAT} result immediately halts progress.

This idea of three output classes is also only a specific instance of a much more general challenge. For arbitrary functions with $n$-bit outputs, predicates represent only $n = 1$. For functions with arbitrary output length $n > 1$, the number of simultaneously satisfiable classes is entirely dependent on function itself; as a result, na\"ive simultaneous maximization fails or reaches computational limits.

\paragraph{A First Step: \texttt{DeriveOutcomes}} Functions with $n$-bit outputs have a maximum of $2^n$ possible output values. However, for many functionalities the actual number of possible outputs is much lower. A natural example is a classification function: each class may be described by a long bit-string (a long output), but only a few classes may exist.

Detecting and exploiting these cases contributes to \mcmpc's practical applicability. At each iteration of an attack, the number of possible outputs defines the number of simultaneous maximization constraints which must be satisfied. Therefore, we provide the \texttt{DeriveOutcomes} subroutine, which uses the \sat~solver to enumerate the possible outputs of the functionality. We perform this step one time at initialization (the \textit{Configure Outcomes} step in Figure~\ref{fig:mcmpc_workflow}), dramatically reducing constraint system size from the $2^n$ maximum in many cases.

\paragraph{Analysis of \texttt{DeriveOutcomes}} \texttt{DeriveOutcomes} is executed once at the beginning of the \mcmpc~workflow. It iterates over all $2^n$ possible outputs of the target $n$-bit output functionality in order to eliminate unreachable outputs. Although this can be concretely expensive, many functions with $n$-bit outputs do not entirely cover their range; eliminating large classes of unreachable outputs dramatically improves performance for the remainder of \mcmpc's steps. Additionally, this step can be skipped, and partial completion still benefits computation time significantly.

\paragraph{Optimizing Leakage: Straw-man Solution} With the set of possible outputs derived, it remains to eliminate large classes of candidates across these output classes at each iteration. Otherwise, the attack is approximately brute-force and therefore highly inefficient. If all $N \leq 2^n$ classes cannot be simultaneously differentiated from one another by a single query, the straightforward next step would be to differentiate \textit{as many classes as possible}.

Here we reach the next challenge: choosing an optimal subset of output classes. The number of subsets of a set of $N$ values is $2^N$, i.e. $2^{2^n}$ for $n$-bit outputs. Even if \texttt{DeriveOutcomes} eliminates many outputs to reduce $N$, the exponential size remains problematic for performance. This combinatorial space is far too large to efficiently iterate through -- worse, doing so would be required at \textit{each iteration} due to evolving constraints. As a result, in order to support non-predicate functions which may contain mutually-exclusive outcome classes, optimality must be sacrificed in favor of an efficient heuristic which performs well in practice.

\paragraph{Solution Intuition} To avoid a combinatorial search, a heuristic must be employed. However, the accuracy of the heuristic directly impacts the profitability of the resulting query, and thus warrants specific analysis and consideration. The intuition for this heuristic (which we refer to as \texttt{SelectOutcomes}) is to remove outcomes from the simultaneous maximization constraint system while minimally affecting profitability.

To achieve this, we perform Model Counting (\sharpsat) on the solution space of each outcome individually. Each result corresponds to the maximum number of candidate models which \textit{might be eliminated} by including that outcome in the simultaneous maximization. We sort all outcomes by their individual sizes, and remove them in ascending order until a simultaneously satisfiability subset is found.

Of course, this approach may miss an ideal configuration of outcomes. However, avoiding combinatorial search is necessary, and in our evaluation (\S\ref{sec:evaluation}), we demonstrate that profitability remains sufficient to discover efficient attacks.

\begin{algorithm}[ht]
    \KwIn{\texttt{Solver}: \sat~solver, \texttt{Pool}: multi-processing pool, $\mathcal{F}$: formula for functionality, $\mathbb{O}$: set of outcomes from \texttt{DeriveOutcomes}}
    \KwOut{$\mathbb{O}^*$: set of mutually-compatible outcomes w/many candidate \texttt{target}, for use in maximization constraints}
        $\mathbb{O}^* \gets \emptyset$\;
        \For{$i\gets1$ \KwTo $|\mathbb{O}|$}{
            $O_i\gets i$th element of $\mathbb{O}$\;
            $\texttt{Solver.constrain}(O_i = \mathcal{F}(\texttt{chosen}, \texttt{target}))$\;
            \tcp{generate CNF for ApproxMC}
            $\phi_i \gets \texttt{Solver.cnf}()$\;
            \tcp{remove constraint for next iteration}
            $\texttt{Solver.pop}()$\;
            \tcp{call \texttt{ApproxMC} in parallel}
            $\texttt{Pool.apply\_async}(\texttt{ApproxMC}, \phi_i)$\;
        }
        
        \tcp{for each outcome and its \texttt{ApproxMC} count}
        \For{$(O_i,cnt)~\textbf{in}~\texttt{Pool.results}()$}{
            \If{$cnt > 0$}{ \tcp{satisfiable single outcome}
            $\mathbb{O}^*\texttt{.add}((O_i,cnt))$\;
            }
        }
        $\mathbb{O}^* \gets$ \textbf{sort} $\mathbb{O}^*$ \textbf{by} $cnt$ \textbf{ascending}\;
        \While{$|\mathbb{O}^*| > 1$ \textbf{and not} \texttt{Solver.satisfiable}$(\mathbb{O}^*)$}{
            $\mathbb{O}^*\texttt{.drop\_first}()$\;
        }
        \tcp{largest simultaneously satisfiable set}
        \textbf{return} $\mathbb{O}^*$\;
\caption{\texttt{SelectOutcomes}}
\label{alg:select_outcomes}
\end{algorithm}

\paragraph{\texttt{SelectOutcomes}} To realize this heuristic, documented in simplified form in Algorithm~\ref{alg:select_outcomes}, we employ a powerful tool from the recent \sat~solving literature: ApproxMC~\cite{approxmc}. ApproxMC (``Approximate Model Counting'') is a tool which takes a CNF Boolean formula and rapidly provides an approximation of the formula's model count. ApproxMC can count complex formulae in a fraction of the time it takes to compute maximization constraints to count formulae. However, it cannot completely supplant maximization: ApproxMC is an external tool which uses sampling to provide approximate counts; there is no known way to efficiently encode iterative sampling  \textit{within the solver} as a constraint, and it is unlikely a method exists due to the data-dependent nature of the ApproxMC sample-and-iterate strategy.

By iterating through the remaining satisfiable outcomes at each iteration of the attack and invoking ApproxMC, we ensure that simultaneous maximization still occurs among as many large classes as can be efficiently identified. Further, we employ process-level parallelism to amortize this sequence of ApproxMC calls. ApproxMC configuration parameters can also trade off single-instance computation time for accuracy if needed. Once a set of mutually-compatible outcomes is found, the attack proceeds with simultaneous maximization using \maxsharpsat. The resulting query is extracted and executed in the to produce a result. By design, this result tends to eliminate a large class of remaining candidate solutions. In the event the result corresponds with a removed outcome class, relatively little knowledge is gained. However, as the removed outcome classes are the smallest, the likelihood that an arbitrary query induces a removed outcome is minimized over a distribution of possible \texttt{target} values.

Algorithm~\ref{alg:select_outcomes} documents the \texttt{SelectOutcomes} heuristic. In the first loop, each outcome is individually constrained to generate a set of formulae using our CNF manipulation interface and the underlying \sat~solver, resetting the solver after each iteration to individually test each outcome. Parallel tasks are dispatched to perform \texttt{ApproxMC} counts of the bits corresponding to the \texttt{target} variable in each formula (bit correspondence requires formula manipulation, omitted for clarity). These results for each outcome are sorted, and until a simultaneously-satisfiable subset is found (more complex than a satisfiability check, but omitted for clarity), the outcomes with the smallest number of candidate \texttt{target} solutions are eliminated. Finally, the usable subset $\mathbb{O}^* \subseteq \mathbb{O}$ is returned.

\paragraph{Analysis of \texttt{SelectOutcomes}} \delphinium~avoids the need for any such heuristics by strictly requiring well-defined and inflexible problem statements. As a result of these strict requirements, the authors of \delphinium~are able to formally prove, probabilistically and approximately, a greedy-optimal algorithm. \mcmpc~enables a far broader scope of functionalities to be analyzed and attacked. Further, reducing the restrictions on the formulae input to \mcmpc~reduces the operator effort and expertise required. By allowing significantly more flexibility in terms of functionality choice and reducing the modeling work required, \mcmpc~sacrifices formal optimality for generality, performance, and practicality.


\subsection{Implementation}\label{sec:impl}

\begin{algorithm}[ht]
    \KwIn{$\mathcal{F}$: Boolean circuit with \texttt{target} and \texttt{chosen} input(s), $\mathcal{O}$: Oracle access to functionality with hidden \texttt{target} input}
    \KwOut{$\mathcal{I}^*$: set of Boolean circuit inputs (vectors of Booleans) for \texttt{chosen} which maximize leakage of \texttt{target} in $\mathcal{F}$}
        $\mathcal{I}^* \gets \emptyset$\;
        $out \gets \texttt{DeriveOutcomes}(\mathcal{F})$\;
        \While{\# of solutions for \texttt{target} > 1}{
            $\overrightarrow{sel} \gets \texttt{SelectOutcomes}(\mathcal{F}, out)$\;
            $query \gets \texttt{Maximize}(\texttt{chosen}, \mathcal{F}, \overrightarrow{sel})$\;
            $result \gets \mathcal{O}(query)$\;
            $\mathcal{F}.\texttt{AddConstraint}(\mathcal{F}(query, \texttt{target}) = result)$\;
            $\mathcal{I}^*.add(query)$\;
        }
        \textbf{return} $\mathcal{I}^*$\;
\caption{\mcmpc~Algorithm Overview}
\label{alg:overview}
\end{algorithm}

\mcmpc~consists of under 2 KLoC of new Python3 which leverages process-level parallelism at every opportunity.
We introduce a CNF translation layer to readily convert \sat~instances and even individual CNF clauses and literals between the two \sat~solvers we employ, CryptoMiniSat and Z3, to reap the relative benefits of each (performance and flexibility, respectively). Our implementation includes a test shim for simulated execution of protocols, and a convenient command-line interface encapsulating numerous configuration options. It is available as open source software on GitHub~\cite{mcmpc_code}. Algorithm~\ref{alg:overview} provides a high-level pseudo-code overview of \mcmpc~using the subroutines described in Section~\ref{sec:beyondpredicates}.

\paragraph{CryptoMiniSat} CryptoMiniSat~\cite{cryptominisat} (CMS) by Soos \textit{et al.} is a \sat~solver designed for cryptographic use-cases. Specifically, CMS includes optimizations for rewriting and processing $Xor$ operations which otherwise incur exponential overhead in the number of operands of a CNF representation. Specifically, an $n$-term $Xor$ expands to $2^{n - 1}$ CNF clauses.

We evaluate \mcmpc~in \S\ref{sec:evaluation} and provide wall-clock computation time to illustrate the practicality of attack generation when using CMS. We confirm the observation of Beck \textit{et al.} that \sat~instances containing $Xor$-dense maximization constraints execute up to an order of magnitude faster in CMS than Z3, and for larger (longer-running) problem instances our parallelized query search offers additional multiplicative factors of time savings.

\paragraph{Z3} Z3~\cite{z3} is an SMT solver which provides extensive and robust software support for Boolean formula manipulation and solving. It is a general-purpose solver which supports arithmetic, bit-vector, and other common theories for ease of use in translating general problems to \sat. We use Z3 for its bit-vector theory support and other tooling. However, Z3 lacks key optimizations which accelerate solving the the particular $Xor$-dense maximization formulae we require. As a result, after formula preparation in Z3, we generate and export a CNF and use CMS for solving, and then recover results back into Z3 for the next round of constraint manipulation.

\section{Evaluation}\label{sec:evaluation}

To evaluate \mcmpc, we assess our implementation against a range of functionalities, from simple motivating examples such as the classic Yao's Millionaires problem~\cite{yao1982protocols} to the recent practical instantiation of MPC developed by researchers at Boston University with the Boston Women's Workforce Council (BWWC) to measure wage equity in a manner which preserved the privacy of participants' salaries~\cite{lapets2016secure}. The success of \mcmpc~in partially or completely deriving confidential inputs across these functionalities demonstrates its usefulness as a tool for practitioners and researchers alike in performing privacy analysis of secure protocols.

\subsection{Selected Functionalities}\label{sec:funcs}

The following selected target functionalities are used to evaluate \mcmpc. Implementations of these functionalities are provided in the open source release using a simple Python3 domain-specific language to describe Boolean formulae. Samples of this DSL can be found in Appendix~\ref{app:dsl}. We provide additional illustrative samples used in our evaluation in our open source release~\cite{mcmpc_code}. These implementations may be useful for further analysis, or to serve as templates for Boolean formula representations of new functionalities.

\renewcommand{\arraystretch}{1.1}
\begin{table}[!t]
    \centering
    \caption{Evaluated Functionalities}
    \label{tab:eval_funcs}
    \begin{tabular}{|l|c|c|}
         \textbf{Functionality} & \texttt{target} & \textbf{Leakage} \\\hline
         \textit{Yao's Millionaires} & $2^{64}$ & $2^{30 \pm 1}$ \\\hline
         \textit{Dual Execution} & $2^{12}$ & $2^{10}-2^{11}$ \\\hline
         \textit{Danish Sugar Beets Auction} & $2^{14}$ -- $2^{56}$ & up to $2^{54}$ \\\hline
         \textit{Bucketed Mean}& $2^{32}$ & $2^{23 \pm 1}$ \\\hline
         \textit{Wage -- Circuit Division} & $2^{36}$ & $2^{34 \pm 1}$ \\\hline
         \textit{Wage -- Standard Division} & $2^{36}$ & $2^{34 \pm 1}$ \\\hline
         \textit{Mean Average} & $2^{8}$ & $2^0 - 2^1$ \\\hline
    \end{tabular}
    \\ \vspace{0.5em} \textit{Refer to \S\ref{sec:funcs}. Smaller domain sizes (\texttt{target}) were chosen to allow many randomized trials. Leakage conservatively estimates eliminated candidates per query, calculated in a single iteration of \mcmpc.}
\end{table}
\renewcommand{\arraystretch}{1}

\paragraph{Evaluated Functionalities} Table~\ref{tab:eval_funcs} lists the evaluated functionalities and the search space size of each corresponding hidden \texttt{target} value. Relatively smaller search spaces were chosen compared with what is practically achievable in reasonable wall-clock time. This allowed evaluation of computation bottlenecks in \mcmpc~while keeping overall evaluation runtime feasible for many randomized repetitions. \mcmpc~outputs estimated leakage per query, useful for evaluating functionalities after only a single iteration of the tool. CNF size growth (in $\#~clauses$) across input sizes is demonstrated in Table~\ref{tab:cnf_sizes} to inform extrapolations.

\subsubsection{Millionaires Problem}
The Millionaires problem introduced by Yao~\cite{yao1982protocols} describes two millionaires, Alice ($A$) and Bob ($B$), who wish to compare their wealth without revealing it. Each has a corresponding input, their net worth represented as 64-bit integers $a$ and $b$, respectively. The functionality to be computed in this simple example is the comparison operation less-than. Thus, at the end of the two-party computation, $A$ and $B$ learn the result of the predicate $a < b$, but learn nothing of each other's input.

This functionality clearly enables a binary search, and the resulting exponential decreases in the search space per query confirm the attack algorithm's approximate optimality (\S\ref{sec:results}). We implement this functionality within \texttt{millionaires.py}. The attack is not particularly subtle, and could certainly be developed and executed manually. However, the example is illustrative, demonstrating the \sat~solver rediscovering the known optimal attack without interaction or guidance.

\begin{table}[!t]
    \centering
    \caption{CNF Size in Clauses over Target Bits}
    \label{tab:cnf_sizes}
    \begin{tabular}{|l|l|l|l|l|}
        \textbf{Functionality} & \textbf{8-bit} & \textbf{16-bit} & \textbf{32-bit} & \textbf{64-bit} \\\hline
         \textit{Yao's Mill.} & 47 & 95 & 191 & 383 \\\hline
         \textit{Dual Exec.} & 875 & 3539 & 14243 & 57155 \\\hline
         \textit{Danish.}$^*$ & 2746 & 6500 & 8701 & 14020 \\\hline
         \textit{BM} & 457 & 561 & 769 & 1185 \\\hline
         \textit{WCD} & ** & ** & 2178 & 5351 \\\hline
         \textit{WSD} & 60 & 749 & 7705 & 42417 \\\hline
         \textit{MA} & 191 & 399 & 815 & 1647 \\\hline
    \end{tabular}
     \\ \vspace{0.5em} \textit{Refer to overview and descriptions in \S\ref{sec:funcs}.}
     \\ \vspace{0.5em} \textit{$^*$Due to encoding, the Danish Sugar Beets functionality was measured at 12, 28, 42, and 60 bits.}
     \\ \vspace{0.5em} \textit{$^{**}$The BWWC WCD function reorganizes division into a multiplication circuit which overflows for small bit-widths.}
\end{table}

\renewcommand{\arraystretch}{1.2}
\begin{table*}[t]
    \centering
    \caption{Evaluation Results}
    \label{tab:results}
    \begin{tabular}{|l|c|c|c|c|c|c|c|}
         \textbf{Functionality} & \texttt{target} & \textbf{Size} & \textbf{Queries} & \textbf{Mean} & \textbf{S.D.} & \textbf{Avg. Time} & \textbf{Parallel Speedup} \\\hline
         Yao's Millionaires & $\fullcirc$ & $2^{64}$ & 69 -- 97 & 84.7 & 8.5 & $\approx$ 3 mins & $\approx 2\times$ \\\hline
         Dual Execution (affine) & $\fullcirc$ & $2^{12}$ & 17 -- 80 & 34.9 & 12.9 & $\approx$ 0.5 mins & $\approx 1.5\times$\\\hline
         Danish Sugar Beets Auction & $\halfcirc$ & $2^{28}$ &  13 -- 13 & 13.0 & 0.0 & $\approx 5$ mins & $\approx 4 - 5\times$ \\\hline
         BWWC Bucketed Mean & $\fullcirc$ & $2^{32}$ & 35 -- 58 & 36.7 & 6.4 & $\approx 6$ mins & $\approx 10\times$ \\\hline
         BWWC Wage (circuit div.) & $\halfcirc$ & $2^{36}$ & 19 -- 26 & 22.3 & 5.0 & $\approx$ 0.5 mins & $\approx 1.5 - 2\times$ \\\hline
         BWWC Wage (standard div.) & $\halfcirc$ & $2^{36}$ & 18 - 29 & 22.4 & 5.8 & $\approx$ 5 mins & $\approx 1.5 - 2\times$ \\\hline
         BWWC Mean Average & $\fullcirc$ & $2^{8}$ & 2 -- 3 & 2.5 & 0.5 & $\approx$ 2.5 mins & $\approx 0.8 - 1.2\times$ \\\hline
    \end{tabular}
    \\ \vspace{0.5em} \texttt{target} uncovered ...partially $\halfcirc$ ...completely $\fullcirc$
    \\ \vspace{0.5em} \textit{Size denotes the size of the \texttt{target} domain. Average time given for full attack, not per-query. Minimum and maximum queries listed with Mean and Standard Deviation. Average queries reported over $\approx 100$ randomized trials. `Speedup' denotes wall clock time savings through parallelization.}
\end{table*}
\renewcommand{\arraystretch}{1.0}

\subsubsection{Dual Execution}
Dual Execution~\cite{mohassel2006efficiency} is an MPC technique which enables conversion of semi-honest secure protocols into malicious secure-with-abort protocols incurring only a single bit of additional leakage. The technique involves both parties in a 2PC garbling a Boolean circuit and sending each other the circuit and necessary data for its evaluation. Both circuits are evaluated, and then a secure comparison protocol informs both parties if the outputs matched. If they do not, the protocol is aborted. The additional bit of leakage comes from the abort, which informs an adversary that the garbled circuit it sent to the honest party did not match the output of the one generated by the honest party -- this knowledge can be leveraged to leak a bit of the honest party's private input upon each execution.

To implement this
part of the adversary's input string is considered to be an encoded program. This program represents the divergent functionality the adversary may choose. For simplicity, we limit the adversary to affine transformations, which can be realized through a matrix multiplication. The complexity of function is entirely up to the adversary, however, allowing configurability is preferable in the dual execution setting where the honest party must be expected to believe they are executing an honest circuit rather than expecting an abort due to a mismatched equality check. The adversary then provides both its own input and the ``code'' the honest party will execute. \mcmpc~is able to uncover all bits of the private input in very few queries (\S\ref{sec:results}), even with the limitation to affine functions.

\subsubsection{Danish Sugar Beets Auction}

The first widely-known practical use of MPC was reported in 2009, when the Danish sugar beets auction was deployed as an MPC. The purpose was to replace the work of a trusted-by-necessity auctioneer with secure computation. The functionality takes in a set of buyer and seller orders to determine the \textit{Market Clearing Price}. An order consists of a set of prices and the number of units (e.g. tons of sugar beets) a buyer/seller is willing to buy/sell at each price. The Market Clearing Price is the equilibrium price at which optimal volumes of sugar beets are exchanged. The functionality is characterized by a matching of buyer and seller prices weighted by the units at each price.

In this protocol, there are a configurable number of buyers and sellers, and we leverage this to test \mcmpc~in the multi-party MPC setting (rather than two-party computation). When modeling adversarial participants, \mcmpc~expresses all adversary inputs as a single vector of Boolean variables over which it determines structure and exploits leakage based on the functionality (Market Clearing Price calculation). This encodes the well-understood behavior of colluding malicious participants in an MPC. We implement this within \texttt{sugarbeets.py}.

\subsubsection{BWWC}

Collaborating with the Boston Women's Workforce Council (BWWC), Lapets \textit{et al.} at Boston University recently developed and deployed an MPC system to detect gender-based wage inequity~\cite{lapets2016secure}. This system allowed companies to privately share aggregate statistics about employee roles, salaries, and gender. The goal was to computationally identify and measure gender-based wage discrimination without requiring any participating companies to directly demonstrate fault. Their design, robustly proven secure under standard assumptions, is an MPC which computes a simple statistical test among each comparable role across companies. By design, their system protects the privacy of each salary data point.

For this target, we developed two aggregation functions inspired by the functionality described. The first was a bucketed mean computation (\texttt{mean\_buckets.py}) which averaged the two inputs and then returned a result representing the number of fixed-size buckets apart the query is from the updated mean. This bucketing procedure implements a non-predicate functionality in which many outcomes are simultaneously satisfiable, exercising the attack algorithm in this regard. The second averages two contributed salary data points (one honest input, one adversary) into a large aggregated salary, and then informs the querying adversary if their contributed salary is below the updated mean. This functionality enables the attack algorithm to derive a binary search for the honest input as demonstrated in the results section (\S\ref{sec:results}).

We developed two forms of this second function, one (\texttt{wage\_circuit\_div.py}) with the division reorganized into a multiplication in the mean computation, and the other (\texttt{wage.py}) with regular integer division inside the solver. Division steps dominate the complexity of this functionality, as demonstrated by the difference between BWWC functions 2 and 3 in Table~\ref{tab:eval_funcs}. Finally, we implemented a plain mean average (\texttt{mean.py}) function (BWWC function 4 in the Table) to evaluate a function with maximal ($2^n$) possible outputs for an $n$-bit \texttt{target}. Despite the small size of the secret in this case, computation time increased due to the large output domain as demonstrated later in this section.

\begin{figure}[!t]
    \centering
    \includegraphics[width=0.95\linewidth]{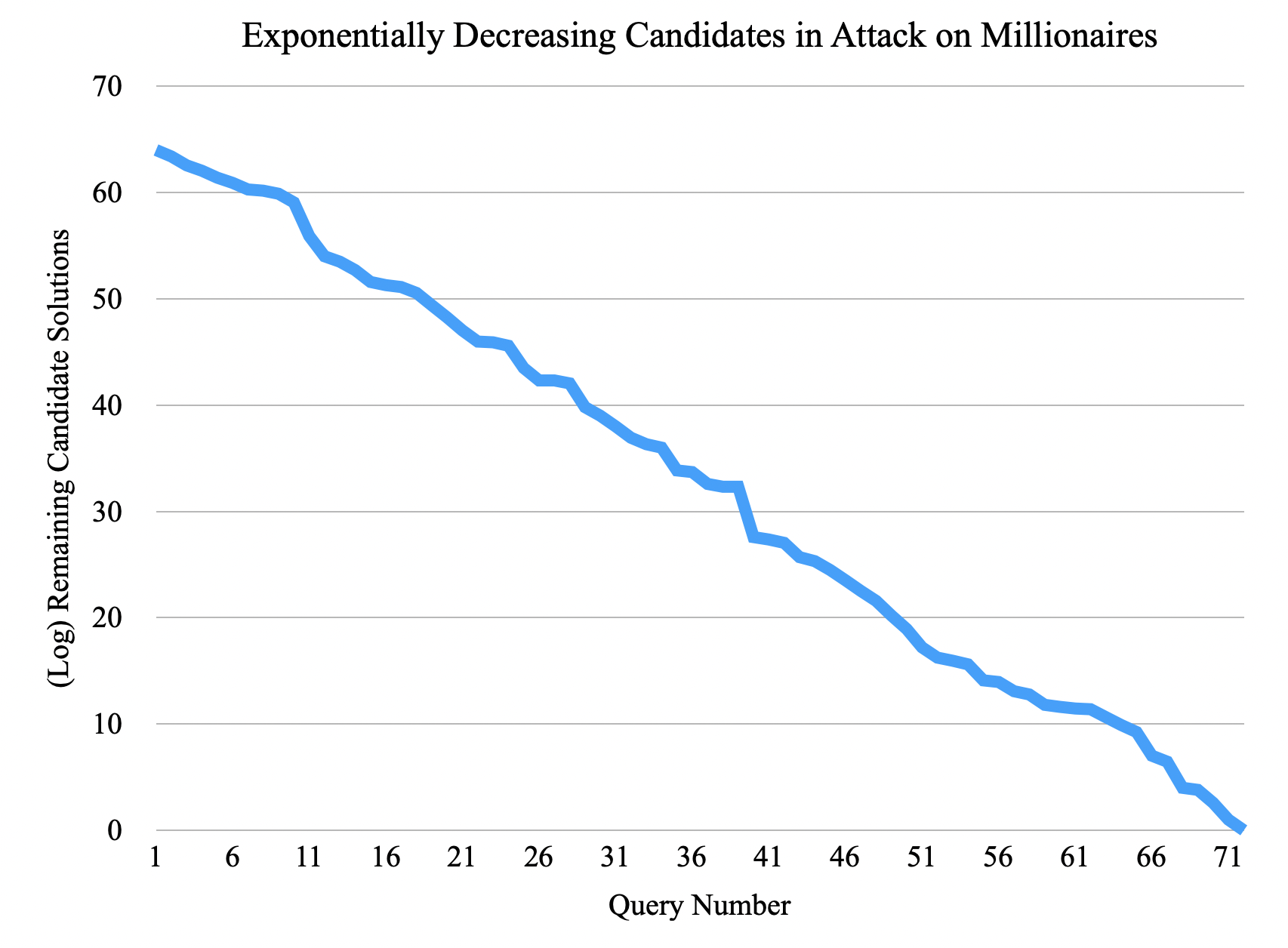}
    \caption{Remaining candidate solutions per query in a single trial of 64-bit Yao's Millionaires}
    \label{fig:cpq_millionaires}
\end{figure}

\subsection{Evaluation Results}\label{sec:results}

In this section, we evaluate \mcmpc~by measuring its effectiveness in uncovering private \texttt{target} input(s) in the selected functionalities. Table~\ref{tab:results} provides a summary of evaluation from repeated trials with random \texttt{target} values.


\paragraph{Process-Level Parallelism} \mcmpc~uses parallelism in two key subroutines, \texttt{SelectOutcomes} and in computing the \texttt{chosen} query at each iteration. For an $n$-bit \texttt{target}, \texttt{SelectOutcomes} spawns one process per possible outcome (up to $2^n$), and computing the query replaces an $O(n)$ linear parameter sweep from \delphinium~with $n$ parallel solving instances. For details on hardware, refer to Appendix~\ref{app:graphs}.

\subsubsection{Millionaires Problem}

As previously discussed, the Yao's Millionaires functionality consists of a comparison between the adversary and hidden inputs. This enables a binary search which, as seen in Figure~\ref{fig:cpq_millionaires}, eliminates approximately half of the remaining candidates at each iteration. The linear descent of the graph on a logarithmic scale captures the efficiency of the attack, resulting in approximately $log(n)$ queries for an $n$-bit \texttt{target}.

Parallelism in the attack generation pipeline achieves an approximately $2\times$ speedup in the Millionaires functionality. Savings occur in the parallel ApproxMC of the two outcomes in the \texttt{SelectOutcomes} heuristic. In this case, however, the heuristic is not required as the comparison predicate is complete and the two outcomes are not mutually exclusive. As a result, \mcmpc~incurs additional overhead in exchange for its ability to handle mutually-exclusive outcome classes. We have added software arguments to configure (optimize) \mcmpc~when the functionality is known to be complete.

\subsubsection{Dual Execution}

In this attack, \mcmpc~generates a sequence of matrices and inputs in order to rule out candidate \texttt{target} inputs. Effectively, this attack creates an increasingly large system of linear equations choosing the coefficients and half the unknowns (\texttt{chosen}) at each step.

\begin{figure}[!t]
    \centering
    \includegraphics[width=0.9\linewidth]{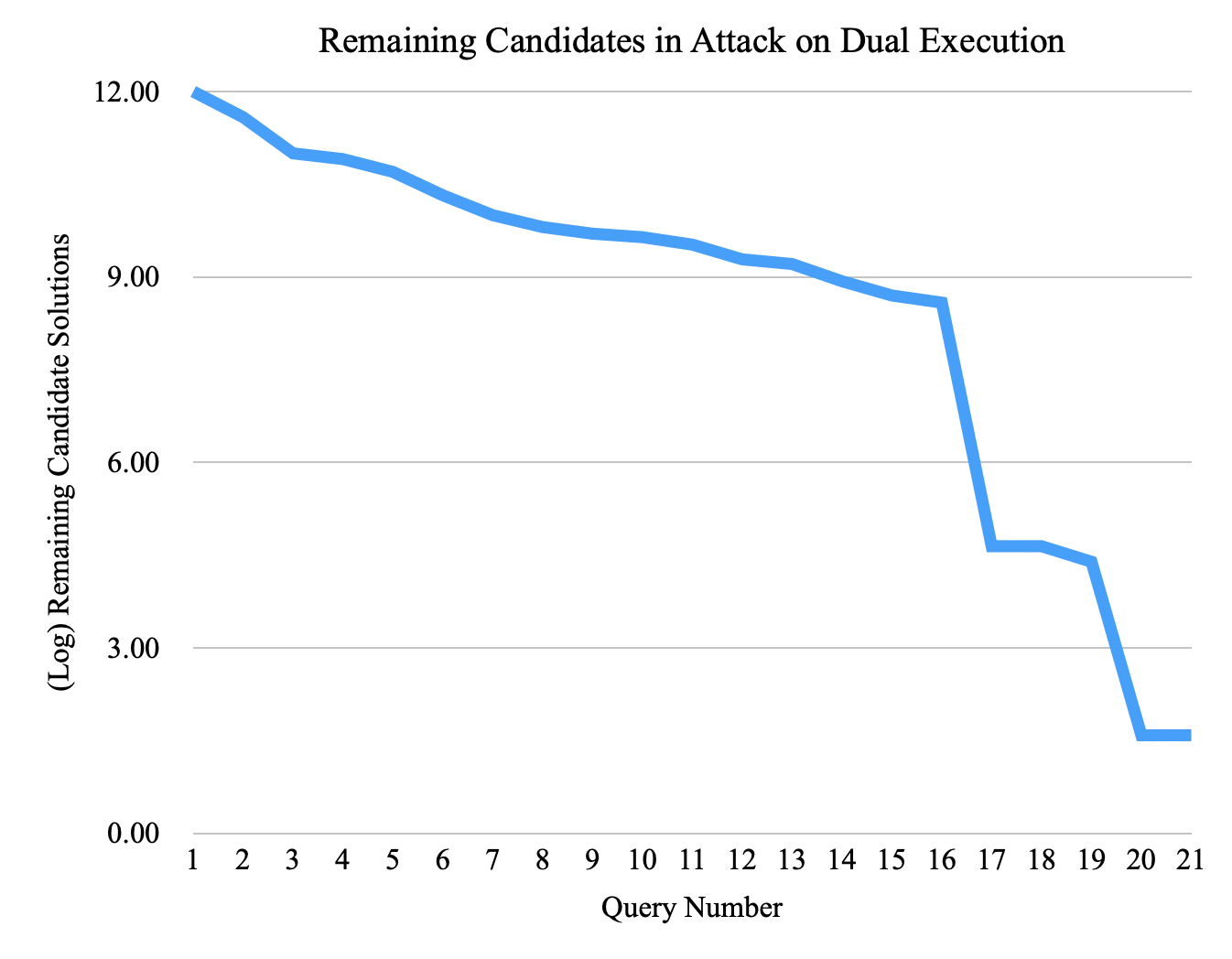}
    \caption{Remaining candidate solutions per query in a single trial of Dual Execution (Affine Predicates)}
    \label{fig:cpq_dualexecution}
\end{figure}

Figure~\ref{fig:cpq_dualexecution} demonstrates the approximately logarithmic decreases of the remaining \texttt{target} search space over the course of an iterative attack. The plateaus visible in the graph further support the idea that there may be points at which \mcmpc~is effectively ``guessing and checking'' within the available information, and once these guesses succeed (after 2-3 attempts, in the visualized attack) the following queries are able to eliminate a large quantity of candidate solutions.

\subsubsection{Danish Sugar Beets Auction}

In the Danish Sugar Beets Auction functionality, not all \texttt{target} bits are uncovered. Figure~\ref{fig:cpq_sugarbeets} in Appendix~\ref{app:graphs} denotes the rapid descent of remaining candidate solutions on a log scale, however, the attack does not reach $2^0 = 1$ (a unique solution). This seems to result purely from insufficient leakage (or, sufficient privacy) of the functionality. At the extremes of low and high prices, \mcmpc~is able to determine the honest buyers/sellers bids for numbers of units. However, at middle prices closer to the likely equilibrium Market Clearing Price, \mcmpc~fails to distinguish and some bits remain unknown.

In this example, we configured the adversary to control 3 buyers and 3 sellers, and the honest party to control 1 buyer and 1 seller. The adversary's goal (and therefore \mcmpc's \texttt{target}) was to uncover the prices and numbers of units of the honest parties' orders. Although only partial knowledge is attained, the security model of the MPC~\cite{bogetoft2009secure} treats the entire prices/amounts list as confidential, and so this attack still represents a meaningful compromise of the intended privacy. Additional parameters and configurations of malicious and honest parties are explored in Appendix~\ref{app:graphs}.

\subsubsection{BWWC} We evaluate three functionalities derived from the Boston Women's Workforce Council MPC~\cite{lapets2016secure}: Bucketed Average, Wage Equity (with two variants), and Mean Average.

\paragraph{Bucketed Average} In the Bucketed Average functionality, \mcmpc~is only able to uncover the \texttt{target} to the granularity of the buckets (ranges). By definition of the functionality, averages which land in the same buckets are indistinguishable, and so this result is expected. Figure~\ref{fig:cpq_mean_buckets} in Appendix~\ref{app:graphs} demonstrates the progress of the iterative attack, noting that it completes before finding a unique solution for \texttt{target}.

Due to the significant number of outcome classes, the \texttt{SelectOutcomes} subroutine dominates execution time. Specifically, the CNF generation step prior to parallelization incurs the most overhead. We discuss the overhead of CNF generation in \S\ref{sec:discussion}. Once CNFs have been exported, parallelized ApproxMC may be employed, and due to the significant number of independent ApproxMC tasks the speedup in this case achieves up to $10\times$ at some iterations.

\paragraph{Wage Equity} In both the \textit{circuit division} and \textit{standard division} variants of the functionality, the wage-average computation leaks significant information of the high-order bits of the honest party's input salary. Although the salary is not uncovered in full, learning the most significant upper half of this value effectively defines the salary range -- disclosing information intended to be private.

In the \textit{circuit division} instance, the mean calculation is reorganized to use a multiplication inside the solver rather than an unsigned integer division. Integer division corresponds with a complex Boolean formula after translation through Z3's SMT interface into \sat~due to the edge-case handling of division. By applying some basic mathematical insight to adapt the target functionality to be more amenable to \sat, we achieve a notable increase in performance.

The performance gain is evident compared to \textit{standard division}. As noted in Table~\ref{tab:results}, the overall computation time at a given bit width differs by an order of magnitude. However, as indicated by the similarity of Figure~\ref{fig:cpq_wage_circuit_div} and Figure~\ref{fig:cpq_wage} in Appendix~\ref{app:graphs}, the resulting attacks achieve similar \textit{per-query} efficiency in eliminating candidate solutions.

Both functionalities encode predicates with two simultaneously satisfiable outcomes. As a result, the \texttt{SelectOutcomes} heuristic is unneeded, and as a result is pure overhead lost in exchange for the assurance that the attack will proceed even if the outcomes are or become mutually incompatible. However, parallelism minimizes the impact of this overhead attaining a $1.5 - 2\times$ speedup across the two outcomes.

\paragraph{Mean Average}
In the Mean Average functionality, the output range of the functionality is the full domain of its inputs. This functionality takes two inputs and computes their mean. As the inputs are bitvectors rather than purely mathematical integers or reals, some complexity is introduced to this averaging through floored division and the $2^n - 1$ bound of an $n$-bit value. As a result, \mcmpc~finds attacks which generally require two queries, occasionally three.

Although this relatively simple functionality admits a query-efficient attack, it is a useful benchmark due to the maximal number of possible outputs. The full-domain output (of exponential size in $n$) taxes \texttt{SelectOutcomes} to the maximum degree per \texttt{target} bit. As a result, the bottleneck of \mcmpc~is the \texttt{SelectOutcomes} heuristic, specifically in CNF generation for each \sat~instance. CNF generation via the Tseitin transformation~\cite{tseitin1968complexity} occurs within the Z3 SMT solver and, while asymptotically efficient, can require significant computation time. Unfortunately, as CNF generation vastly exceeds ApproxMC solving time within \texttt{SelectOutcomes}, process-parallelism offers little benefit and even occurs slight overhead in some tests.

Mean Average-like functionalities are the key motivator for the \texttt{SelectOutcomes} heuristic. With a \delphinium-like approach, the attack immediately fails with an UNSAT result. The reason for this failure becomes clear by example.

Consider a 2-bit adversary input $a$, and 2-bit honest input $b$. $\texttt{mean}(a, b) \in \{0, 1, 2, 3\}$. Configuring simultaneous maximization to find an optimal $a$, the following constraint (among others) is added to the solver. The symbolic representations of the four partitions of $b$ denoted $\{b_0, b_1, b_2, b_3\}$ correspond to $b$ when the mean with $a$ is $0$, $1$, $2$, and $3$, respectively.
\begin{align*}
mean(a, b_0) = 0~\land~mean(a, b_1) = 1~\land\\
mean(a, b_2) = 2~\land~mean(a, b_3) = 3
\end{align*}

\noindent Notice that the first clause of the conjunction requires $(a, b_0)$ to be $(0, 0)$, $(0, 1)$, or $(1, 0)$ (with floored division). Therefore $a \in \{0, 1\}$. However, the last clause requires $(a, b_3) = (3, 3)$ exclusively. As $\{0, 1\} \cap \{3\} = \emptyset$, the conjunction is unsatisfiable. \texttt{SelectOutcomes} correctly identifies that $b_1$ and $b_2$ are not similarly mutually exclusive, and that they contain the largest number of candidate solutions for $b$. As a result, $b_0$ and $b_3$ are selected out, and the attack proceeds without interruption. \mcmpc~is then able to derive an optimal or near-optimal attack in 2 -- 3 queries.


\subsection{Contributed Benchmarks}\label{sec:benchmarks}



As an additional artifact, we have submitted a wide assortment of benchmarks to the SMT-LIB compendium of \sat~benchmarks~\cite{BarFT-SMTLIB}. These have been accepted to help aid solver development and increase performance on the particular instance types \mcmpc~encounters. Advancements in \sat~research can offer a drop-in improvement to this work by expanding the horizon of computational feasibility. A summary of the submitted files is presented in Appendix~\ref{app:bench}. A subset of these benchmarks have already begun to see use within \sat~research as a basis to test a new model counting approach~\cite{arjit}.

\section{Discussion}\label{sec:discussion}

Our results demonstrate that \mcmpc~is capable of evaluating a diverse array of simple functionalities in a matter of minutes of computation time. The generated attacks, which rely on the aforementioned multi-run adaptive assumption, are able to uncover all bits of the private \texttt{target} value in many cases. When our approach cannot continue to differentiate classes of candidate solutions to eliminate, it notifies the user that the attack may devolve to ``brute-force'' and asks if they'd like to continue. This occurs when \texttt{SelectOutcomes} fails to find a simultaneously satisfiable set of outcomes.

We observe that predicate functionalities execute relatively quickly, and functionalities with larger outputs (e.g. the Danish sugar beets auction and bucketed mean functionalities) spend significant time in \texttt{SelectOutcomes}. On the extreme end, the mean average functionality has an output domain as large as the output bitvector allows -- and we correspondingly observe the most time spent in navigating these highly mutually-exclusive outcome classes.

In multiple cases, the \texttt{SelectOutcomes} heuristic is the computational bottleneck for attack progression. Analyzing these cases more closely reveals that the CNF generation step, using Tseitin's method~\cite{tseitin1968complexity} to generate a CNF from an arbitrary Boolean formula with only linear expansion, is the crux of the subroutine. Despite its asymptotic efficiency, this step (as executed within Z3) takes significant time. We employ a number of CNF manipulation and caching techniques to avoid unnecessary regeneration of CNFs where possible, however, the relative speed of parallelized ApproxMC leaves this sequentially-executed step as the longest-running component.

Even without the multi-run adaptive assumption to allow an iterative attack, the initial analysis step quantifies approximately how much leakage can be exploited with a given optimized query. On its own, this enables privacy/confidentiality analysis of any functionality planned for inclusion into an MPC, FHE, or ZK scheme. Further, this analysis needs to be run only once for a given functionality, and so even if hours or days of computation are required for some complex functionality, the resulting quantified leakage or iterative attack can provide pivotal insights to researchers and developers.

\paragraph{Limitations} The remaining practical limitations of \mcmpc~ are largely entangled with the inherent complexity of the computational problems it employs. For sufficiently complex functionalities or large output domains, running time increases significantly. Depending on the use-case, this may not rule out the use of \mcmpc~for important/sensitive functionalities or contexts. Notably, for functionalities with very large output domains, or which encode cryptographic primitives (e.g. the AES round function~\cite{aes-key-recovery}) directly within their Boolean circuitry, \mcmpc~reaches wall-clock performance bottlenecks which may impede its applicability.

\mcmpc~seeks to reduce operator burden and required expertise to develop confidentiality attacks or measure leakage. However, a certain degree of operator involvement is still required: \mcmpc~can only be as accurate as the Boolean formula representation of a target functionality.
The Python3 DSL we provide is an initial step in this direction, however it still requires understanding and manual effort.





\section{Conclusion}\label{sec:conclusion}

\sat~and cryptography have intersected numerous times in the research literature, often to their mutual benefit. \mcmpc~pursues this interdisciplinary exchange by applying emerging \sat~techniques to a new domain, bridging theory and practice, and contributing back to the \sat~community. With \mcmpc, developers of secure protocols can automatically determine privacy thresholds or generate attacks against candidate systems. Potential users of these tools can evaluate them before choosing to use them. For sufficiently sensitive use cases, extensive computation times for complex or large functionalities may be worthwhile; \mcmpc~empowers practitioners to make that decision without requiring expensive expert analysis.

\section*{Acknowledgments}
We would like to thank the anonymous reviewers and Shepherd for their helpful comments and feedback. This work was supported in part by National Science Foundation (NSF) under grants CNS-20-46361, a DARPA Young Faculty Award (YFA) under Grant Agreement D22AP00137-00, and by DARPA under Contract No. HR001120C0084. Any opinions, findings and conclusions or recommendations expressed in this material are those of the authors and do not necessarily reflect the views of the United States Government, the NSF, or DARPA. The authors would also like to thank Katarina Mayer for her contributions to this work.

\bibliographystyle{plain}
\bibliography{main}

\onecolumn
\begin{appendix}

\section{Solver DSL Samples}\label{app:dsl}

Python3 DSL example for Yao's Millionaires:
\begin{lstlisting}
def func_smt(solver, chosen_input, target_input):
    """ Millionaire's functionality inside the solver """
    return solver._if(solver._ugt(chosen_input, target_input), # Millionaire's
                      solver.bvconst(1,OUTCOME_LEN),
                      solver.bvconst(0,OUTCOME_LEN))

def func(chosen_input, target_input):
    """ Millionaire's functionality outside the solver """
    return chosen_input > target_input
\end{lstlisting}

\noindent Python3 DSL example for Dual Execution (affine predicates):
\begin{lstlisting}

def func_smt(solver, chosen_input, target_input):
    # isolate adversary-chosen function (matrix) from adversary-chosen input
    matrix_bits = solver.extract(chosen_input, CHOSEN_LEN-1, TARGET_LEN)
    chosen_input = solver.extract(chosen_input, TARGET_LEN-1, 0)
    # unpack matrix values
    matrix = [[solver.extract(matrix_bits, j*TARGET_LEN+i, j*TARGET_LEN+i)
               for i in range(TARGET_LEN)] for j in range(TARGET_LEN)]
    chosen_bits = [solver.extract(chosen_input, i, i) for i in range(TARGET_LEN)]
    target_bits = [solver.extract(target_input, i, i) for i in range(TARGET_LEN)]
    # perform mults
    chosen_matrix = [reduce(lambda x, y: solver._add(x, y),
                            [solver._mult(matrix[j][i], chosen_bits[i])
                             for i in range(TARGET_LEN)])
                     for j in range(TARGET_LEN)]
    target_matrix = [reduce(lambda x, y: solver._add(x, y),
                            [solver._mult(matrix[j][i], target_bits[i])
                             for i in range(TARGET_LEN)])
                     for j in range(TARGET_LEN)]
    # re-pack matrices
    chosen_out = solver.concat(*reversed(chosen_matrix))
    target_out = solver.concat(*reversed(target_matrix))
    # evaluate equality check
    return solver._if(solver._eq(chosen_out, target_out),
                      solver.bvconst(1,1),
                      solver.bvconst(0,1))

def func(chosen_input, target_input):
    matrix_bits = chosen_input >> TARGET_LEN
    chosen_input = chosen_input & ((1 << TARGET_LEN)-1)
    matrix = [[(matrix_bits >> (i+TARGET_LEN*j)) & 1 for i in range(TARGET_LEN)]
              for j in range(TARGET_LEN)]
    chosen_bits = [(chosen_input >> i) & 1 for i in range(TARGET_LEN)]
    target_bits = [(target_input >> i) & 1 for i in range(TARGET_LEN)]
    chosen_matrix = [sum([(matrix[j][i]*chosen_bits[i])%2 for i in range(TARGET_LEN)])
                        % 2
                     for j in range(TARGET_LEN)]
    target_matrix = [sum([(matrix[j][i]*target_bits[i])%2 for i in range(TARGET_LEN)])
                        % 2
                     for j in range(TARGET_LEN)]
    chosen_out = 0
    for i, bit in enumerate(chosen_matrix):
        chosen_out |= bit << i
    target_out = 0
    for i, bit in enumerate(target_matrix):
        target_out |= bit << i
    return 1 if chosen_out == target_out else 0
\end{lstlisting}

\section{Additional Evaluation}\label{app:graphs}
\begin{figure*}[t]
    \centering
    \begin{subfigure}[b]{0.48\textwidth}
    \centering
    \includegraphics[width=\textwidth]{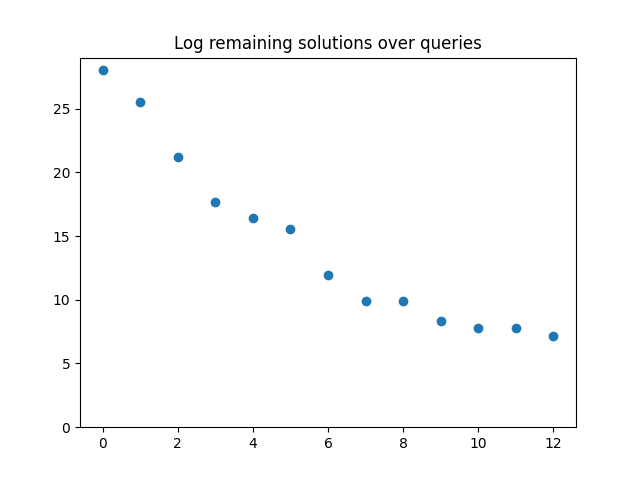}

    \caption{Danish Sugar Beets Auction ($2^{28}$)}
    \label{fig:cpq_sugarbeets}
\end{subfigure}
\hfill
\begin{subfigure}[b]{0.48\textwidth}
    \centering
    \includegraphics[width=\textwidth]{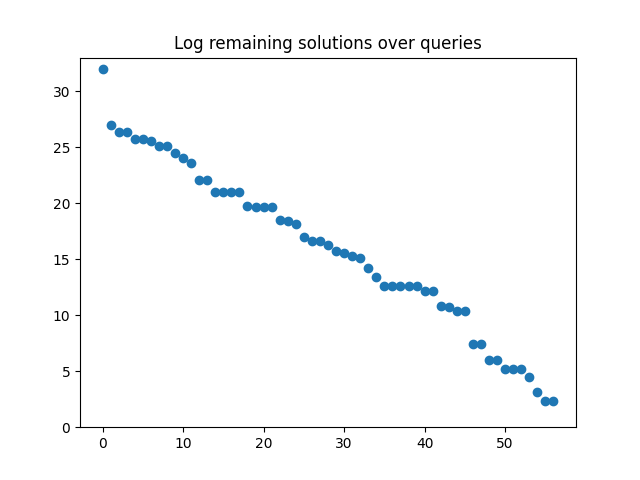}
    \caption{BWWC Bucketed Mean ($2^{32}$)}
    \label{fig:cpq_mean_buckets}
\end{subfigure}

\begin{subfigure}[b]{0.48\textwidth}
    \centering
    \includegraphics[width=\textwidth]{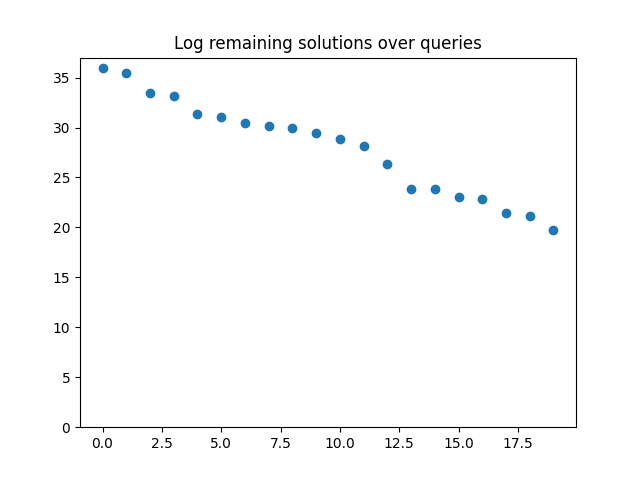}
    \caption{BWWC Wage (circuit division) ($2^{36}$)}
    \label{fig:cpq_wage_circuit_div}
\end{subfigure}
\hfill
\begin{subfigure}[b]{0.48\textwidth}
    \centering
    \includegraphics[width=\textwidth]{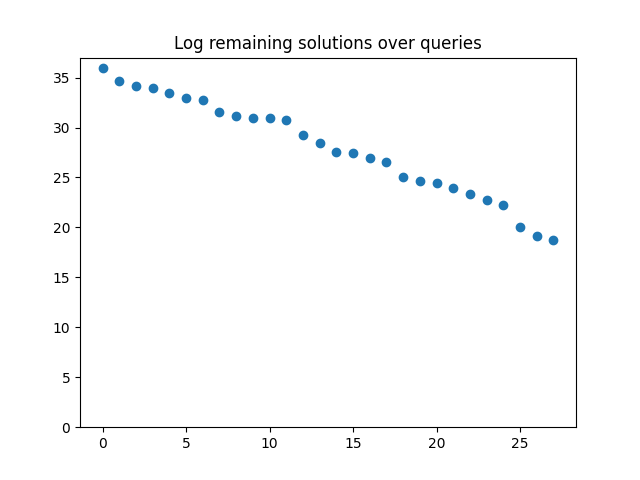}
    \caption{BWWC Wage (standard division) ($2^{36}$)}
    \label{fig:cpq_wage}
\end{subfigure}
    \caption{Remaining candidate solutions (log scale) per query}\label{fig:extra_graphs}
\end{figure*}
\noindent Each graph within Figure~\ref{fig:extra_graphs} below denotes a representative example attack generated by \mcmpc. Each test operates over a domain of 28- to 36-bit values in order to keep overall benchmarking time reasonable with many repetitions of each attack. Some attacks do not trend to $2^0$ as the functionality does not admit complete leakage of the underlying secret (honest) input(s).

\paragraph{Testing Environment} All evaluations were performed on an Intel Xeon E5 CPU at 2.10GHz with 500GB RAM. Process-level parallelism was configured to employ 64 of the available virtual threads of execution using a Python3 standard library process pool~\cite{python3_pool}. We used CryptoMiniSat 5.8.0, ApproxMC 4.0.1, Z3 4.8.15, and Python3 3.8.10.

\paragraph{Many-Party MPC in the Sugar Beets Auction} The Sugar Beets Auction functionality demonstrates \mcmpc~in the setting where more than two MPC participants are involved. In effect, \mcmpc~treats multi-party protocols as 2PC: the solver is simply aware of bits that are symbolic and out of its control, and bits that it is able to vary to induce satisfiability. However, as it is an illustrative example, we demonstrate the varying results of \mcmpc~as the number of honest and colluding malicious parties varies. Often, the uncovered \texttt{target} bits does not completely cover the domain, but in these cases \mcmpc~generally uncovers one or more high bits of each Honest party's input, restricting their possible values to smaller ranges within the domain.

\renewcommand{\arraystretch}{1.2}
\begin{table*}[t]
    \centering
    \caption{Additional Sugar Beets Auction Evaluation}
    \label{tab:auction_results}
    \begin{tabular}{|l|c|c|c|c|c|}
         \textbf{Functionality} & \texttt{target} & \textbf{Domain Size} & \textbf{Queries} & \textbf{Avg. Time} & \textbf{Parallel Speedup} \\\hline
         4 Players -- All Sellers Malicious & 14 -- 18 & $2^{28}$ &  11 -- 15 & $\approx 5$ mins & $\approx 4 - 5\times$ \\\hline
             4 Players -- All Buyers Malicious & 8 -- 18 & $2^{28}$ &  8 -- 15 & $\approx 5$ mins & $\approx 4 - 5\times$ \\\hline
         4 Players -- Half Each Malicious & 12 -- 12 & $2^{28}$ & 11 -- 18 & $\approx 5$ mins & $\approx 4 - 5\times$ \\\hline
         4 Players -- One Buyer Malicious & 0 & $2^{42}$ & 6 -- 10 & $\approx 3$ mins & $\approx 4 - 5\times$ \\\hline
         4 Players -- Three Buyers Malicious & 14 & $2^{14}$ & 8 -- 9 & $\approx 1$ mins & $\approx 2\times$ \\\hline
         6 Players -- 2 Malicious 1 Honest Each & 10 -- 14 & $2^{28}$ & 11 -- 12  & $\approx 5$ mins & $\approx 2 - 4\times$ \\\hline
         6 Players -- 1 Malicious 2 Honest Each & 8 -- 20 & $2^{56}$ & 12 -- 20  & $\approx 20 - 30$ mins & $\approx 1 - 7\times$ \\\hline
    \end{tabular}
    \\ \vspace{0.5em} \texttt{target} bits discovered (maximum $log_2(domain)$)
    \\ \vspace{0.5em} \textit{Average time given for full attack, not per-query. Minimum and maximum queries listed.\\ `Speedup' denotes wall clock time savings through parallelization.}
\end{table*}
\renewcommand{\arraystretch}{1.0}

\section{Contributed Benchmarks}\label{app:bench}

Throughout the evaluation of \mcmpc, numerous complex \sat~formulae were generated and stored as text files. The \sat~community aggregates such files to serve as benchmarks and testing tools for international competitions of \sat~solver speed and capability. Table~\ref{tab:benchmarks} summarizes the benchmark CNF files derived from each functionality and characterizes them in terms of clause and variable counts given as ranges and an average across all files. These CNFs have been accepted to the SMT-LIB benchmark collection~\cite{BarFT-SMTLIB} used in \sat~solver competitions which motivate research and improvement. Each CNF instance is paired with an equivalent SMT2 file defined in the quantifier-free bitvector (\texttt{QF\_BV}) domain; these equivalent but differently encoded files have proven useful in \sat~research alone and in contrast to their CNF pairs~\cite{arjit}. Finally, in the course of generating, gathering, and testing numerous randomized \sat~instances, a variety of underlying software bugs in the CryptoMinisat~\cite{cryptominisat} and Z3~\cite{z3} solvers were uncovered, reported to open-source software maintainers, and resolved collaboratively.

\renewcommand{\arraystretch}{1.5}
\begin{table}[ht]
    \centering
    \caption{Benchmarks Overview}
    \label{tab:benchmarks}
    \begin{tabular}{l|c|c|c|c|c}
        \textbf{Functionality} & \textbf{Instances} & \textbf{CNF Clauses} & \textbf{Avg. Clauses} & \textbf{CNF Variables} & \textbf{Avg. Variables} \\\hline\hline
        \textbf{Yao's Millionaires} & 1,504 & 1,158 -- 62,676 & 17,116 & 506 -- 17,039 & 4,524 \\\hline
        \textbf{Dual Execution (affine)} & 729 & 4,898 -- 49,160 & 8,562 & 1,549 -- 12,375 & 2,290 \\\hline
        \textbf{Danish Sugar Beets Auction} & 88 & 49,020 -- 159,349 & 112,178 & 10,451 -- 27,640 & 19,834 \\\hline
        \textbf{BWWC Bucketed Mean} & 345 & 29,794 -- 66,138 & 46,238 & 8,824 -- 17,036 & 12,593 \\\hline
        \textbf{BWWC Wage (circuit div.)} & 194 & 6,058 -- 11,053 & 7,655 & 1,283 -- 2,465 & 1,571 \\\hline
        \textbf{BWWC Wage (standard div.)} & 210 & 21,898 -- 94,822 & 75,813 & 4,987 -- 12,208 & 9,072 \\\hline
        \textbf{BWWC Mean Average} & 28 & 1,646 -- 76,440 & 49,966 & 503 -- 19,682 & 13,408\\\hline\hline
        \textbf{Total} & 3,098
    \end{tabular}
\end{table}
\renewcommand{\arraystretch}{1}

\section{Zero-Knowledge Range Proofs}

\paragraph{Note on ZK Range Proofs} A range proof~\cite{bunz2018bulletproofs} is a computational proof that a value lies within a given range. Range proofs can be performed in zero-knowledge protocols, meaning that the verifier learns nothing of the value being tested, only the result of the computation. Such zero-knowledge proofs are useful in anonymous payment systems which require aggregate payment validation without betraying information about individual payments to the broader payment network~\cite{sasson2014zerocash}. These zero-knowledge proofs can be considered a 2PC between a prover (holding a secret value) and a verifier (holding a range to be tested against the value) wherein the functionality executed is the aforementioned range validation check and the privacy of the prover's value must be maintained. The logic of a range proof effectively matches that of the Millionaires problem, and so this target is included simply to highlight \mcmpc's capacity to analyze a ZK protocol. We offer our software implementation as a tool to researchers and developers of novel ZK protocols as a method to evaluate privacy loss.

\end{appendix}

\end{document}